# Modelling Complexity: the case of Climate Science

Valerio Lucarini [v.lucarini@reading.ac.uk]
Department of Meteorology
Department of Mathematics
University of Reading, Reading, UK

*...Chaos is the future
and beyond it is freedom
confusion is next and next after that is the truth..*[1]

*...llud in his quoque te rebus cognoscere avemus,
corpora cum deorsum rectum per inane feruntur
ponderibus propriis, incerto tempore ferme
incertisque locis spatio depellere paulum,
tantum quod momen mutatum dicere possis.
Quod nisi declinare solerent, omnia deorsum,
imbris uti guttae, caderent per inane profundum,
nec foret offensus natus nec plaga creata
principiis: ita nil umquam natura creasset...*[2]

*Abstract*
*We briefly review some of the scientific challenges and epistemological issues related to climate science. We discuss the formulation and testing of theories and numerical models, which, given the presence of unavoidable uncertainties in observational data, the non-repeatability of world-experiments, and the fact that relevant processes occur in a large variety of spatial and temporal scales, require a rather different approach than in other scientific contexts. A brief discussion of the intrinsic limitations of geo-engineering solutions to global warming is presented, and a framework of investigation based upon non-equilibrium thermodynamics is proposed. We also critically discuss recently proposed perspectives of development of climate science based purely upon massive use of supercomputer and centralized planning of scientific priorities.*

### 1. Introduction

The climatic system is constituted by four intimately interconnected sub-systems, atmosphere, hydrosphere, cryosphere, and biosphere, which evolve under the action of macroscopic driving and modulating agents, such as solar heating, Earth's rotation and gravitation (Peixoto and Oort 1992). The climate system features many degrees of freedom - which makes it *complicated* – and nonlinear interactions taking place on a vast range of time-space scales accompanying sensitive dependence on the initial conditions – which makes it *complex*. In Table 1 we present some simple examples aimed at clarifying the difference between complex and complicated systems. The distinction between these two concepts is further clarified by considering the origin of the two words: "complex" comes from the past participle of the Latin verb *complector, -ari* (to entwine), whereas "complicated" comes from the past participle of the Latin verb *complico, -are* (to put together) (Lucarini 2002).

---
[1] Moore T, Confusion is next, in Sonic Youth's album *Confusion is sex*, distributed by Neutral (1983)
[2] Lucretius, De Rerum Natura, II, 216-224

Table 1: Complex vs Complicated systems. Examples from natural sciences.

|  | Not Complex | Complex |
|---|---|---|
| Not Complicated | Harmonic oscillator | Lorenz 63 model |
| Complicated | Gas of non-interacting oscillators (e.g. phonons) | Turbulent fluid |

The description of the macroscopic dynamics of the climate system is based on the systematic use of dominant *balances* derived on a *phenomenological* basis in order to *specialize* the dynamical equations. Such balances are suitable classes of approximate solutions of the evolution equations representing a reasonably good approximation to the actual observed fields when sufficiently large spatial or temporal averages are considered (Speranza and Lucarini 2005). Actually, different balances have to be considered depending on the time and space scales we are focusing our interest on. Depending on the time scale of interest and on the problem under investigation, the relevant *active* degrees of freedom (mathematically corresponding to the separation between the *slow* and *fast* manifolds), needing the most careful representation, change dramatically. For relatively short time scales (below 10 years) the atmospheric degrees of freedom are active while the other sub-systems can be considered essentially *frozen*. For longer time scales (100-1000 years) the ocean dominates the dynamics of climate, while for even longer time scales (over 5000 years) the continental ice sheet changes are the most relevant factors of variability (Saltzman 2002).

Such an approach reflects the fundamentally heuristic-inductive nature of the scientific research in this field, where the traditional reductionist scientific approach is not necessarily effective. Climate science is a quickly evolving subject resulting from the intersection of growing number of disciplines, such as:
- Meteorology, Oceanography, Remote Sensing, Radiative Transfer;
- Statistical Physics, Thermodynamics, Fluid Dynamics;
- Chaotic and Stochastic Dynamical Systems;
- Statistics, Data Assimilation, Data reconstruction from Proxy indicators;
- Numerical Methods in Modelling;
- Biology, Ecology, Geochemistry.

In recent years, several authors have attempted the systematization of the growing body of research dealing with complex systems under the label of *Complexity*. Numerous books and journals are being published under this, rather successful, brand, and it is encouraging to see a ever increasing degree of collaboration and exchange between social and natural scientists. In Fig. 1 we report an example of a map of complexity, which tries to underline that the degree of interconnection between different subfields is such that the unique scientific framework of complexity can be defined. Interestingly, most if not all of the proposed maps of complexity feature a notable absence, precisely that of climate science.

One could propose that in the map presented in Fig. 1, the large empty space in the upper right corner should be filled by a balloon referring to climate science. Such an absence is considerably puzzling if one considers that some of the most notable features shared by most complex systems (e.g. sensitive dependence on initial conditions, multiscale properties, intermittency) have been discovered in the context of or in vicinity to climate problems, and that climate science, especially in the last two decades, has emerged to being one of the most widely discussed scientific fields. Maybe, this is actually the reason of such a notable absence: climate science is perceived as a mostly policy-driven, high-tech, computer muscled-up field, rather than being a fronteer subject where to test and improve the refined tools and concepts needed to analyze and deal with complexity.

Therefore, it is clear that the investigation of the global structural properties plays a central role for the provision of a unifying picture of the climate system. Such an endeavour is of fundamental importance for improving substantially our understanding climate variability and climate change on a large variety of scales, which encompass major paleoclimatic shifts, almost regularly repeated events such as ice ages, as well as the ongoing and future anthropogenic climate change, as envisioned by the scientific programme proposed in the landmark book by Saltzman (2002).

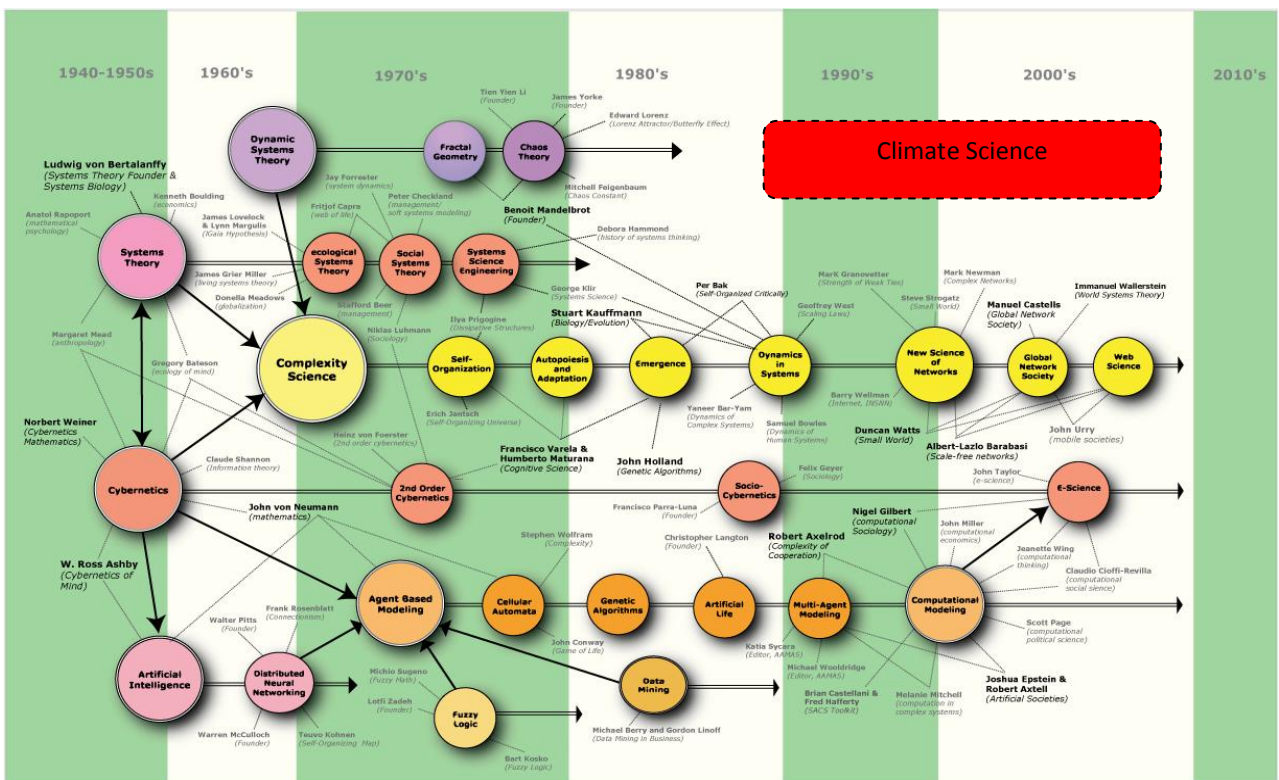

**Figure 1:** Adapted from the map of complexity proposed by B. Castellano (http://www.personal.kent.edu/~bcastel3/). The balloon indicating Climate Science has been added by the author.

Such an effort has significant relevance also in the context of the ever-increasing attention paid by the scientific community to the quest for validating climate models (CMs) of various degrees of complexity, as explicitly requested by the Intergovernmental Panel on Climate Change (IPCC) in its 4[th] Assessment Report (IPCC 2007), and for the definition of strategies aimed at the radical improvement of their performance (Held 2005; Lucarini 2008a).

In a modern, global perspective, the climate can be seen as a complex, non-equilibrium system, which generates entropy by irreversible processes, transforms moist static energy into mechanical energy as if it were a heat engine, and, when the external and internal parameters have fixed values, achieves a steady state by balancing the input and output of energy and entropy with the surrounding environment (Peixoto and Oort 1992, Johnson 2000, Lorenz and Kleidon 2005, Lucarini 2009a). The tools of phenomenological non-equilibrium thermodynamics (de Groot and Mazur 1962) seem very well suited in defining a new point of view for the analysis of the CS for understanding its variability and its large-scale processes, including the atmosphere-ocean coupling, the hydrological cycle, as well as understanding the mechanisms involved in *climate phase transitions* observed at the so-called tipping points (Lenton et al. 2008), i.e. conditions under which catastrophes may occur for small variations in the boundary conditions or in the internal parameters of the system (Fraedrich 1979).

Moreover, a primary goal of climate science is to understand how the statistical properties – mean values, fluctuations, and higher order moments - of the climate system change as a result of modulations to the parameters of the system occurring of various time scales. A large class of problems fall into this category, such as those involving climate sensitivity, climate variability, climate change, climate tipping points, as well as the response to daily, seasonal, orbital forcings, to changes in the atmospheric composition, to changes in the geography and topography of the continents and of the seafloor. Recent results from non-equilibrium statistical mechanics mostly due to Ruelle (1998, 2009) provide rigorous tools for tackling this problem using a perturbative approach (Lucarini 2008b, 2009b; Lucarini and Sarno 2010).

2. **Issues in Climate Modelling**

Given the nature of their research, numerical simulation has been a key method of investigation for climate scientists since the early days of computers. Actually, in the late 1940s, the first large-scale application of automatic computing consisted in the first numerical weather forecast, based on greatly simplified equations, which was proposed by Von Neumann and mainly devised by Charney. This also emphasizes the long-standing strategic relevance of climate-related science. Since the late 1950s, the US (and Swedish) technical services have been using computer-assisted numerical integration of relatively accurate equations descriptive of the physics of the atmosphere to routinely produce weather forecasts.

The evaluation of the accuracy of numerical climate models and the definition of strategies for their improvement are, today more than ever, crucial issues in the climate scientific community. On one side climate models of various degrees of complexity constitute tools of fundamental importance to reconstruct and project in the future the state of the planet and to test theories related to basic geophysical fluid dynamical properties of the atmosphere and of the ocean as well as of the physical and chemical feedbacks within the various subdomain and between them. On the other side, the outputs of climate models, and especially future climate projections, are gaining an ever increasing relevance in several fields, such as ecology, economics, engineering, energy, architecture, as well as for the process of policy-making at national and international level. Regarding influences at societal level of climate-related finding, the impacts of the IPCC (2007) report are unprecedented, to the point that the Panel was awarded the 2007 Nobel Prize for Peace.

Numerical modelling options strongly rely on the available computer power, so that the continuous improvements in both software and hardware have permitted a large increase in the performances of the models and at the same time an impressive widening of their horizons. On one side, the adoption of finer and finer resolutions has allowed a more detailed description of the large scale features of the dynamics, and, more critically, a more direct physical description of a larger set of processes, thus limiting the need for parameterisation procedures, which, where needed, have become more accurate. On the other side, it has been possible to implement and then refine the coupling between models pertaining to different systems having a common boundary, such as the atmosphere and the ocean, or the atmosphere and the land surface.

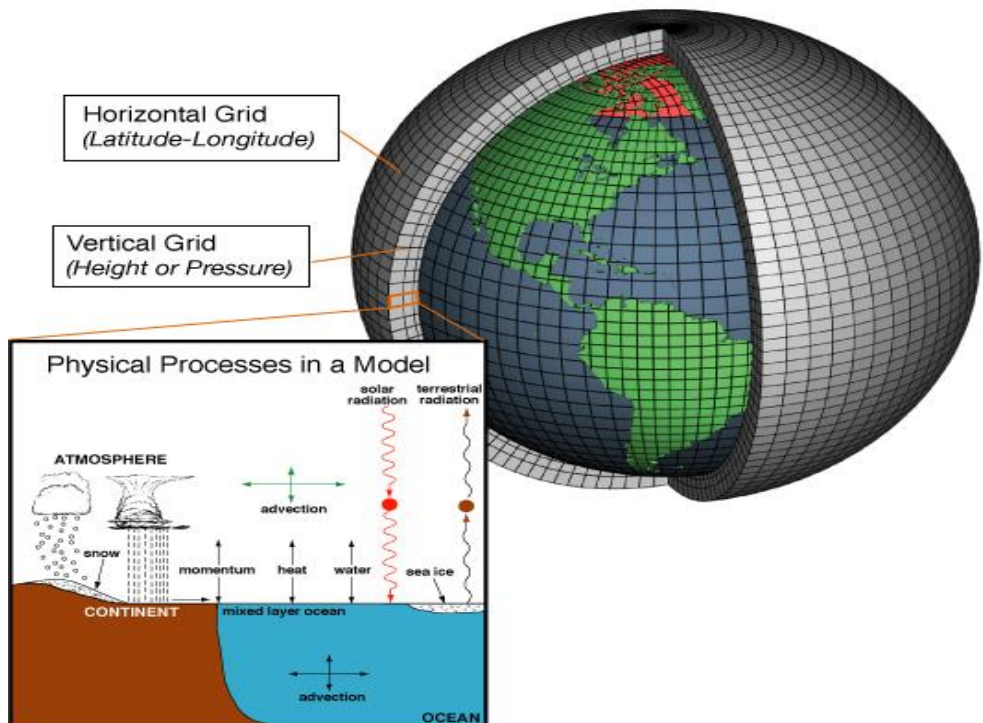

**Figure 2: Overview of the structure of a state-of-the-art climate model. From the NOAA website http://www.research.noaa.gov/climate/t_modeling.html**

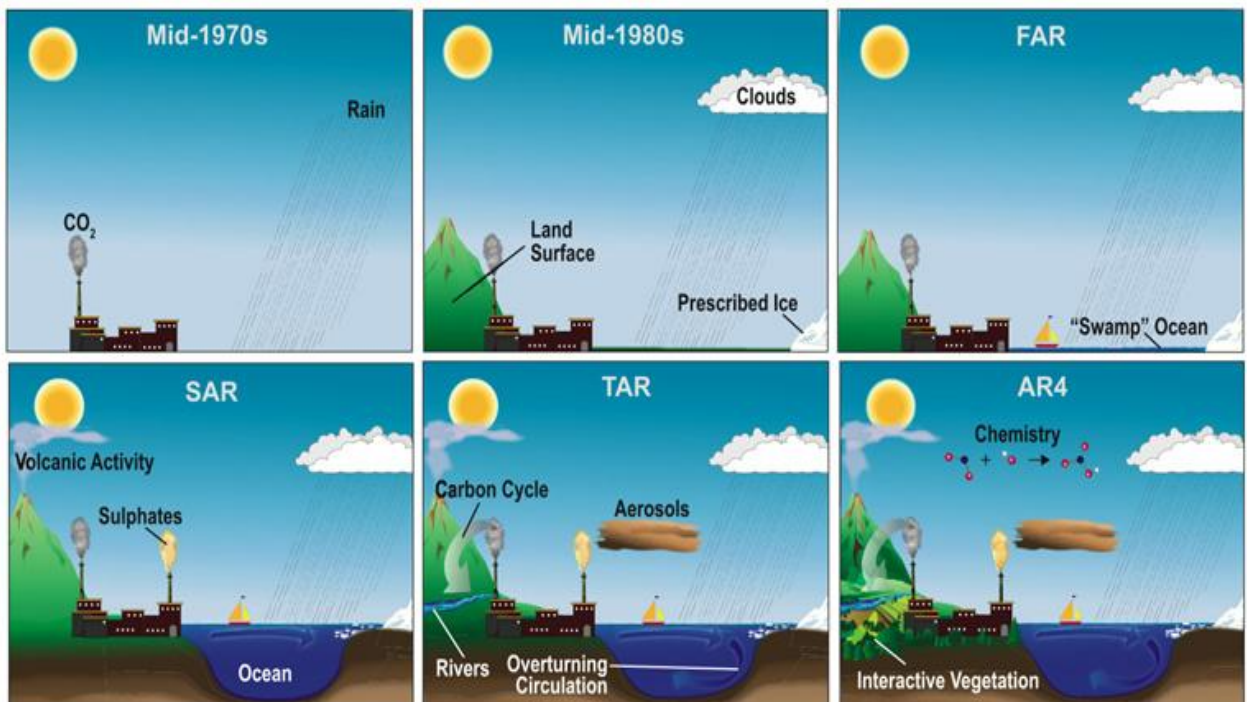

**Figure 3: Evolution of state-of-the-art Climate Models from the mid 70s to the mid 00s. From IPCC (2007)**

Still, since the climate is a multiscale system (Schertzer and Lovejoy 2004), our ability to represent it with numerical models is intrinsically limited. One should consider that climate variability is observed between spatial scales of $10^{-6}$ m to $10^7$ m and between $10^{-6}$ s and $10^{16}$ s, range that dwarf what covered explicitly by present top-notch models by many orders of magnitude. The progress in terms of computing power of a factor of $10^6$ obtained in the last 30 years has reduced only by a relatively small amount the distance between model and the actual system, so that it seems unfeasible to expect within the next decades fundamental progresses to our understanding of the climate system obtained only through brute force computing. This is in sheer contrast with what envisioned by Navarra et al. (2010).

Climate modelling faces uncertainties belonging to two distinct classes. The uncertainties on the initial conditions (*uncertainties of the first kind*) limit, because of the chaotic nature of the system, our ability to predict deterministically the state of the system at a future time given our imperfect knowledge of its state at the present time. In the growing body of research dealing with climate prediction, this kind of uncertainty is partially taken care of by applying the same strategies today commonly adopted in the usual weather forecasting, i.e. by using ensemble simulations. Along these lines, many simulations are started with slightly perturbed initial conditions, and the set of evolved trajectories is used to provide a probabilistic estimate of how the system will actually evolve. Obvious limitations are related to the technological difficulties of running a sufficient number of ensemble members. But more basic problems are also present. The structural deficiencies together with an unavoidably limited knowledge of the external forcings (*uncertainties of the second kind*) limit intrinsically the possibility of providing realistic simulations of the statistical properties of the climate system, especially affecting the possibility of representing abrupt climate change processes.

The validation, or auditing – overall evaluation of accuracy - of a set of climate models, is a delicate operation, which can be decomposed in two related, albeit distinct, procedures. The first procedure is the intercomparison, which aims at assessing the consistency of the models in the simulation of certain physical phenomena over a certain time frame. The second procedure is the verification, whose goal is to compare the models outputs to corresponding observed or reconstructed quantities. A third kind of uncertainty is related to the actual procedure of auditing: what are the best *metrics,* i.e. the best statistical estimators to be used for analysing the output of climate models? In principle, any reasonable function of the variables included in our climate model is a perfectly legitimate metrics. Nevertheless, even is all such *observables* are mathematically well-defined, their physical relevance and robustness can be very different. Since no strict a-priori criterion exists for selecting a good observable, even if taking into account some basic physical properties of the climate system can provide useful guidance, as explained below, we do not have a unique recipe for testing our models. Again, this is in sheer contrast with the case of more traditional scientific fields, where the relevant observables (e.g., in high-energy physics, "mass", "transition probability", "cross-section") are suggested by the very equations we are trying to solve or analyse experimentally.

3. **Performance Metrics and Uncertainties**

A matter of great interest in the analysis of climate models is the choice of the physical observables used in the auditing procedures, or, as they are often referred to, of the metrics of validation of the climate models. An ever-increasing attention paid by the scientific community to the quest for reliable, robust metrics, as explicitly requested by the $4^{th}$ Assessment Report of the IPCC.

Most typically, the models' validation is based upon the analysis of the skill in simulating fields of common practical interest, such as the surface air temperature or the accumulated precipitation. However, these fields describe quantities that can hardly be considered *climate state variables*. By considering the vertical profile of the annual and global mean temperature, the zonal mean surface air temperature or precipitation, the impression is that all models have very similar performances and it is very difficult to assess whether a model is performing in any sense better than any other. Nevertheless, they differ substantially in the horizontal as well as vertical resolution, numerical schemes, physical parameterisations, and so on.

One aim – from the end-user's point of view – is immediately checking how *realistic* the modelled fields of practical interest are. But if the aim is to define strategies aimed at the radical improvement of their performance, beyond incremental advances often obtained at the price of large increases in requested computed power, it is important to fully understand the differences in the representation of the *climatic machine* among models and possibly decide whether specific physical processes are correctly simulated by a specific model.

In order to analyse the representation of specific physical processes as well as of balances involving conservation principles, it is necessary to use specialized diagnostic tools - that we may call *process-oriented metrics* - as indexes for model reliability. Such approach may be helpful in clarifying the distinction between the performance of the models in reproducing *diagnostic* and *prognostic* variables of the climate system. The definition of efficient process-oriented metrics benefits from the adoption of a well-defined scientific framework. In section 6 we propose our point of view where we maintain that a thermodynamic perspective is well-suited for analysing the climate system, because it provides a way to cut through its complexity and, at the same time, carefully take into account its non-equilibrium properties.

Additional practical as well as epistemological issues emerge when we consider the actual process of comparing theoretical and numerical investigations with experimental data. Model results and approximate theories can often be tested only against observational data from the past, which may feature problems of various degrees of criticality, essentially because of the physical extension of the systems under analysis. The available historical observations sometimes feature a relatively low degree of reciprocal synchronic coherence and individually present problems of diachronic coherence, due to changes in the strategies of data gathering with time, whereas proxy data, by definition, provide only semi-quantitative information on the past state of the climate system. The natural variability of both the model and of the real system contributes to blur the line between a failed and a passed test. Anyway, a positive result would not at all ensure the model's ability to provide consistent future projections, whereas at most it is possible to deduce out of a negative result that the model is not reliable enough. Summarizing, difficulties basically emerge because we always have to deal with three different kinds of attractors: the attractor of the real climate system, its reconstruction from observations and the attractors of the climate models.

The unavoidable presence of such critical uncertainties implies that every model used to generate projections about future climate change could be interpreted as being weak in its descriptive power. Climate science does not have real or virtual laboratories where theories and models can be tested against experiments, since phenomena often take place only once, due to the entropic time arrow, and cannot be reproduced. Using a standard scientfiic procedure, if a model fails to comply with even just one observable, it should be rejected. That's how, e.g., high energy physics typically works, as shown by the very idea of building the Large Hadron Collider (LHC). As clear from the previous discussion, it is unfeasible to use this criterion in climate science, because we would end up discarding all models and arresting any progress. Therefore, the Galilean scientific framework given by recurrent interplay of experimental results and theoretical predictions is challenged.

As for taking care of possible issues related to initial conditions, often an ensemble of simulations, where the same climate model is run under identical conditions from slightly different initial state; this allows a more detailed exploration of the phase space of the system, with a better sampling – on a finite time - of the attractor of the model.

The deficiencies of a single climate model and the stability of its statistical properties can be addressed, by applying Monte Carlo techniques to generate an ensemble of simulations, each characterized by different values of some key uncertain parameters characterizing the global climatic properties. Therefore, in this case sampling is performed by considering attractors that are parametrically deformed, which is, by the way, a formally well-defined operation when we consider the Ruelle (1998, 2009) response theory (see discussion in Lucarini 2008b).

A detailed analysis of structural uncertainties requires the comparison of different models following a horizontal and vertical conceptual hierarchical path. The horizontal comparison is the comparative study of the results generated by models sharing a roughly common level of complexity, but having been implemented in different ways by different people. The vertical comparison is the comparative study of the results obtained by a family of models, each built as an extension and complexification of another one starting from an initial simple parent, thus creating a natural hierarchy of increasing complexity.

The Project for Climate Model Diagnostics and Intercomparison (PCMDI), through its climate models intercomparison projects (CMIPs), has supported the gathering into a single web-location of climate model outputs contributing to the activities initiated by the IPCC. The PCMDI thus provides a unique opportunity for evaluating the state-of-the-art capabilities in simulating the behaviour of climate system. The CMIP has provided a rather complete and standardized set of climate outputs in its third phase, which was related to the IPCC (2007) report, whereas the CMIP's fifth phase will collect data relevant for the preparation of the fifth assessment report of IPCC.

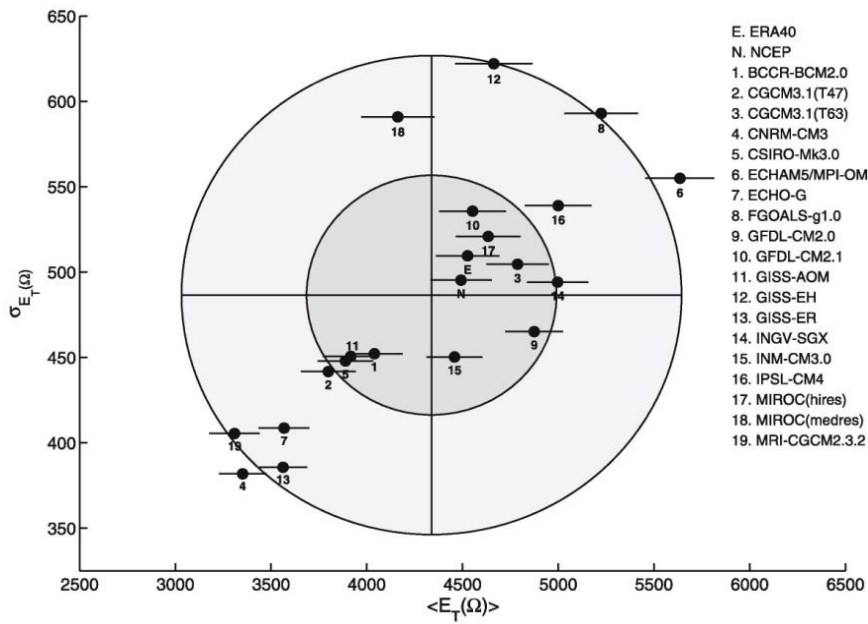

**Figure 4: Performance of various state-of-the-art climate models in representing winter mid-latitude northern hemisphere atmospheric variability. Climatology of integrated spectral power of the waves is plotted against its interannual variability (in $m^2s^{-2}$). The various points correspond to the dataset indicated in the legend. The ensemble mean is located at the center of the ellispses. Adapted from Lucarini et al. (2007).**

In order to describe synthetically and comprehensively the outputs of a growing number of climate models, recently it has become common to consider multi-model ensembles and focus the attention of the ensemble mean and the ensemble spread of the models, taken respectively as the (possibly weighted) first two moments of the models outputs for the considered metric. Then, information from rather different attractors is merged. Whereas this procedure has surely advantages, such statistical estimators should not be interpreted in the standard way - the mean approximating the truth, the standard deviation describing the uncertainty – because such a straightforward perspective relies on the (false) assumption that the set is a probabilistic ensemble, formed by equivalent realizations of given process, and that the underlying probability distribution is unimodal. Figure 4 portraits the statistical properties of the (x-axis: mean value; y-axis: interannual variability) of a quadratic measure of the strength of winter northern hemisphere mid-latitude atmospheric disturbances during the period 1961-2000 and reports the results for 19 state-of-the-art climate models included in the PCMDI dataset. Moreover, reference data are reported for the two reanalyses datasets, produced by NCEP-NCAR and ECMWF, commonly considered as roughly equivalent reconstructions to the true atmospheric state. As we see, the ensemble mean (center of the two ellipses) is actually rather close to the "true" state, but, on the other hand, it is positioned in a location where the density of the points referring to the outputs of the various models is very low. Note that the two semi-axes of the internal (external) ellipsis are given by (twice) the values of the standard deviation of the ensemble for the two considered variables. Therefore, it is at least questionable to interpret the ensemble mean as representative in any well-defined sense of the models' outputs.

### 4. A side note: an outlook on Geoengineering

In spite of all the efforts of several scientific communities, pressure groups, and citizens, the more and more widespread concern regarding the climatic impacts of the observed steady increase of CO2 concentration in the atmosphere has not been met with the actual provision of an effective, international and multilateral protocol of economic and political measures aimed at limiting present and future hazards. Being able to deal effectively, and in the context of an increasingly multipolar world, with the complexities of the global economy and politics and of the global climate system at the same time seems an almost insurmountable task. In this context, in recent years a growing number of scientists, policy-makers, and corporations have proposed the adoption of geo-engineering strategies as – at least – short term mitigation of climate change effects due to CO2 increase.

In general, geo-engineering refers to the adoption of measures aimed at modifying, on purpose, the climate system in a – allegedly – controlled way. On smaller temporal and spatial scales, several weather modification strategies have been devised in the course of the years, such as the seeding of clouds aimed at increasing their rain efficiency. Nevertheless, geo-engineering is distinct as its scope is intrinsically global in space and multiannual in time. One of the most relevant proposal in this direction is considered has been that of continuously injecting in the atmosphere large amounts of aerosols in order to reduce the amount of net incoming solar radiation (some aerosols reflect quite efficiently the solar radiation), thus countering the anthropogenic greenhouse effect due to ever-increasing $CO_2$ concentration. This idea has been evaluated as technologically feasible and economically very convenient with respect to challenging the present model of economic development. Putting aside the ethical issues related to the idea of countering pollution with further pollution and those to the fact that a single country or, in principle, even a private can decide to alter unilaterally the global climate, and neglecting the large scientific uncertainties still surrounding the actual effects of such large-scale injection of material in the atmosphere, the complexity of the climate system seem to suggest that this kind of operation is intrinsically ill-posed, or better, far from being a simple solution to a complex problem like global warming is.

Mathematically, we can say that geo-engineering is about defining suitable isolines constructed in the following way. If we consider an increment x of $CO_2$ concentration, what is the amount of aerosols y needed to keep constant the average value of the statistical properties of the climate variable z? By changing the value of x and finding the corresponding values of $y=y_z(x)$, we construct the isoline of the climate variable z, i.e., when moving parametrically along such a line (corresponding to the adoption of geo-engineering measures contrasting the increase in $CO_2$ concentration), the climatology of z is not altered. But, if we choose any other climate variable $z_1$, $z_2$, .., $z_n$, the geo-engineering strategy will not provide any solution, because $z_1$, $z_2$, .., $z_n$ are, instead, constant along the isolines $y=y_{z1}(x)$, ..., $y=y_{zn}(x)$, which are in general be distinct from $y=y_z(x)$. Therefore, along the parametric curve $y=y_z(x)$ the value of the climate variables $z_1$, $z_2$, .., $z_n$ will definitely change, so that *climate will change*. Injecting the aerosols in the atmosphere has the effect of modulating, but not of erasing in any real sense, the effect of increasing CO2 concentrations.

Therefore, the geo-engineering strategy described by $y=y_z(x)$ will only provide an example of constrained climate change scenario, and not at all a scenario foreseeing the cancellation of climate change in general. We then understand that all the emphasis is in the selection of the z-variable of interest, and it seems rather clear that such a choice has an eminently political nature, and, furthermore, it seems hopeless to reach a global consensus on the "right" variable to consider in a hypothetically pro-geo-engineering world. In Fig. 5 we provide a graphical representation of this issue, where we consider four variables (globally averaged surface temperature ($\langle T_S \rangle$), averaged sea surface temperature ($\langle T_{SS} \rangle$), averaged atmospheric temperature ($\langle T_A \rangle$) and surface temperature averaged over the land located in the mid-latitudes of the northern hemisphere ($\langle T_{S,NHML} \rangle$). Each country or group of countries will have different and even contrasting interests, as the effects of climate change are felt locally and the adaptive capacity are widely different. Note that, by definition, potential strategies aimed at achieving a reduction of the CO2 concentration in the atmosphere do not suffer from this problem. Therefore, geo-engineering seems to be a logical loop-hole, as rather than providing a practical solution to the ongoing anthropogenic forcing, moves the difficulties to the choice.

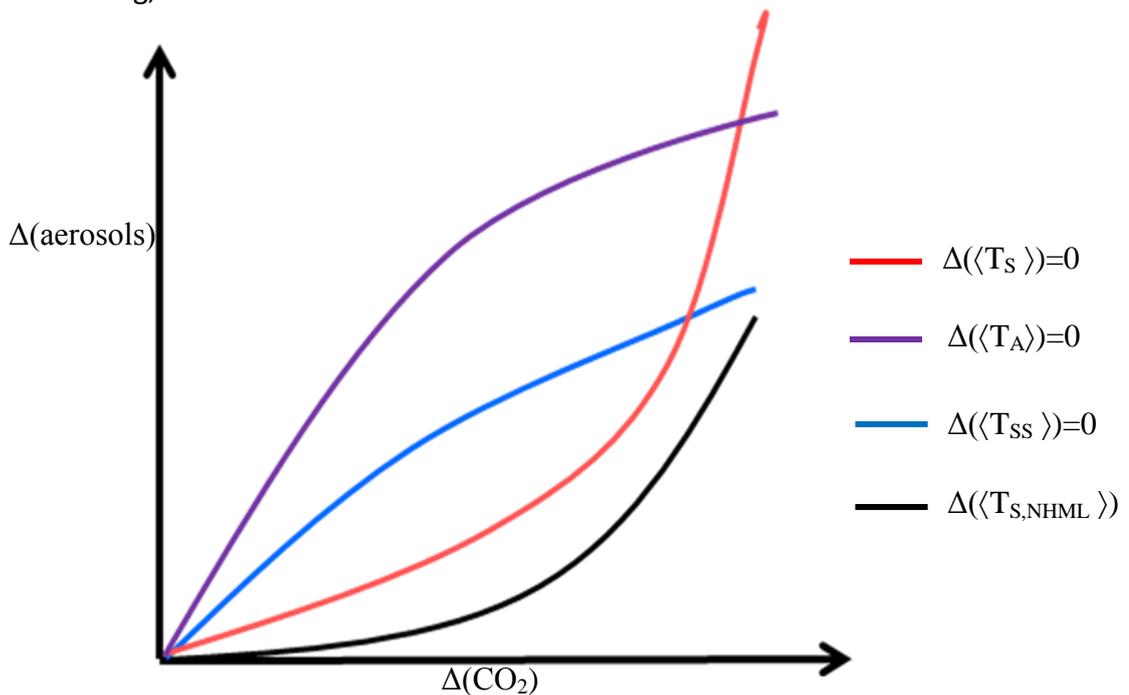

**Figure 5:** Apart from the scientific uncertainties, geo-engineering measures provide a fix only for a selected climate observable. The isolines of $\langle T_S \rangle$, $\langle T_A \rangle$, $\langle T_{SS} \rangle$, and $\langle T_{S,NHML} \rangle$ are, in fact different. See details in the text.

## 5. Our proposal: a Thermodynamic perspective

Many authors have approached the problem of understanding the properties of the CS by studying the structure of the bifurcations of dynamical systems constructed heuristically and featuring a minimal number of climatically relevant variables (usually below 10). This strategy has brought to great scientific results and has been suggested the existence of generic mathematical structures, sometimes re-discovered in hierarchies of CMs. A relevant example of investigation performed along these lines on processes occurring on multi-decadal time scale is the analysis of the stability of the thermohaline circulation. On atmospheric time scales, some of the most important investigations of the low-frequency variability of the mid-latitude atmosphere have been carried out along similar lines. The limitations of this approach lie on the fact that the simplifications adopted in the derivation of the dynamical systems may blur out the involved physical processes and hardly allow for an efficient representation of the fluctuations of the system, to which the introduction of stochastic forcing provides a partial solution (Hasselmann 1976). This approach suffers from need for a - usually beyond reach - closure theory for the noise properties.

While acknowledging the scientific achievements obtained along the above mentioned line, we propose a different approach for addressing the *big picture* of a complex system like climate is. An alternative way for providing a new, satisfactory theory of climate dynamics able to tackle simultaneously balances of physical quantities and dynamical instabilities is to adopt a thermodynamic perspective, along the lines proposed by Lorenz (1967). We consider simultaneously two closely related approaches, a phenomenological outlook based on the macroscopic theory of non-equilibrium thermodynamics (see e.g., de Groot and Mazur 1962), and, a more fundamental outlook, based on the paradigm of ergodic theory (Eckmann and Ruelle 1985) and more recent developments of the non-equilibrium statistical mechanics (Ruelle 1998, 2009).

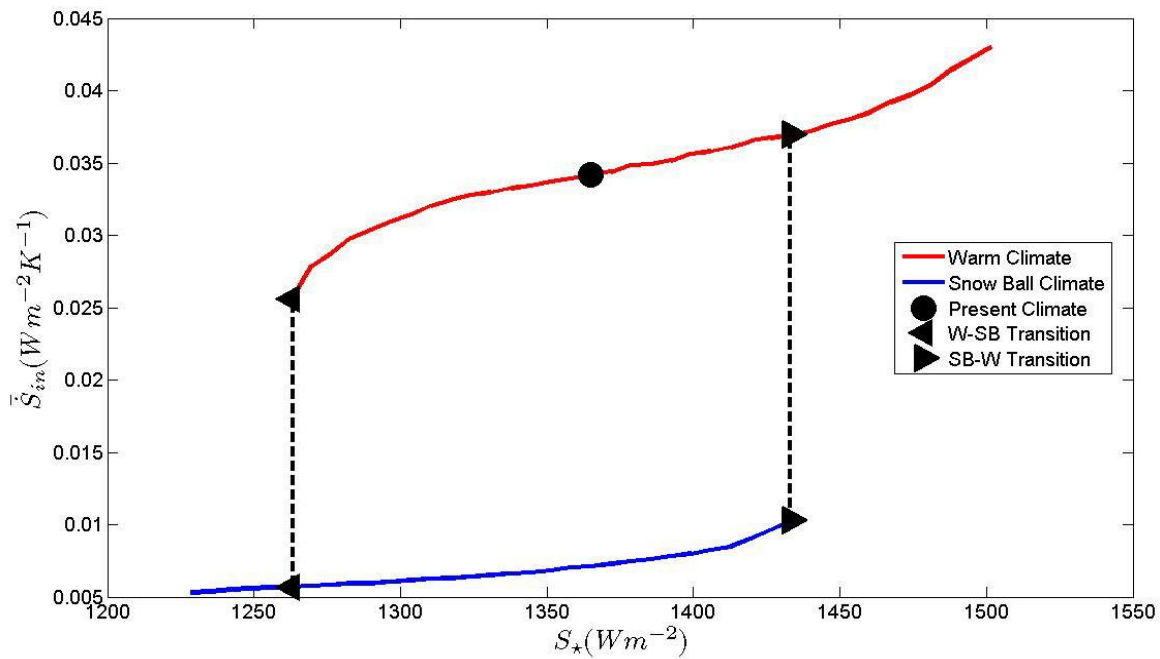

**Figure 6: Dependence of the Entropy Production of the climate system on the value of the solar constant. Note the presence of a wide region of bistability, where both the warm (W) and the snowball (SB) climates are stable. Adapted from Lucarini et al (2010ba).**

The concept of the energy cycle of the atmosphere introduced by Lorenz (1967) allowed for defining an effective climate machine such that the atmospheric and oceanic motions simultaneously result from the mechanical work (then dissipated in a turbulent cascade) produced by the engine, and re-equilibrate the energy balance of the climate system. One of the fundamental reasons why a comprehensive understanding of climate dynamics is hard to achieve lies on the presence of such a nonlinear closure. Recently, Johnson (2000) introduced a Carnot engine–equivalent picture of the climate system by defining effective warm and the cold reservoirs and their temperatures. The interest towards the studying the climate irreversibility largely stemmed from the proposal of the maximum entropy production principle (MEPP), which suggests that a non-equilibrium nonlinear systems adjust so to maximize the entropy production (Ozawa et al 2003, Kleidon and Lorenz 2005). Even if recent claims of *ab initio* derivation of MEPP have been dismissed, it has stimulated the re-examination of entropy production in the climate system (Pascale et al. 2009) and the development of new strategies for improving the CMs parameterisation.

Recently, a link has been proposed between the Carnot efficiency, the intensity of the Lorenz energy cycle, the entropy production and the degree of irreversibility of the climate system (Lucarini 2009a). In particular, it has been found that the efficiency of the equivalent thermal machine sets also the proportionality between the internal entropy fluctuation of the system and the lower bound to entropy production by the fluid compatible with the $2^{nd}$ law of thermodynamics. Such a bound is basically given by the entropy produced by the dissipation of the mechanical energy, whereas the excess of entropy production is due to the transport of heat down the gradient of the temperature field. These results pave the way for a new, extensive exploration aimed at understanding the climate response under various scenarios of forcings, of atmospheric composition, and of boundary conditions. Recent preliminary efforts have focused on the impacts on the thermodynamics of the climate system of changes in the solar constant, with the analysis of the onset and decay of snowball Earth conditions (Lucarini et al., 2010a), and on those due to changing $CO_2$ concentration (Lucarini et al., 2010b). In the snowball Earth experiment, the two climate regimes (ice-covered and today-like) feature radically different physical properties. In particular, the climate efficiency decreases (increases) with increasing solar constant in present (snowball) climate conditions. Moreover, entropy production (see Fig. 6) and the irreversibility of the system are much higher in warmer climates. When considering in $CO_2$ changes, a warmer CS results to be less efficient, more irreversible, and produces more entropy. While in cold climates a dominating role for the changes in the thermodynamics is played by changes in the meridional albedo (fraction of solar radiation scattered back to space) contrast, in warm ones changes in latent heat fluxes are crucial. Many interesting questions remain to be addressed.

As the results in Lucarini (2009) allow for treating the exchange of mechanical energy between atmosphere and ocean as a boundary term in the energy budget, this approach may contribute to quantifying the mechanisms involved in the mechanical energy budget in the global ocean, which have long been source of debate in oceanography (e.g. Wunsch and Ferrari 2004, Tailleux 2010). Additionally, a thermodynamic analysis of the climate transitions at the tipping points (Lenton et al. 2008) based upon macro-scale thermodynamic properties is also proposed. In (Lucarini et al. 2010a) it is shown that the loss of stability of a climate regime is accompanied by the transition to a regime featuring a less efficient climate, which is characterized by thermodynamic conditions closer to equilibrium. It seems very relevant to tackle the analysis of the suggestive hypothesis of the generality of this behaviour. This has implication for the issue of multiple stability in the atmosphere-biosphere system.

Recently, it has been shown that it is possible to compute the entropy production and derive information on the Lorenz energy cycle by only looking at the 2D fields of top-of-the-atmosphere and surface radiative budgets (Lucarini et al. 2010c). This paves the way for studying the thermodynamics of the *climates* of planetary bodies other than the Earth, whose investigation has been, buy the way, one of the first application of MEPP (Kleidon and Lorenz 2005). This is a rather promising perspective, given the ever increasing attention paid to, and data obtained on, these astronomical objects.

The fundamental approach based upon non-equilibrium statistical mechanics provides great opportunities, due the recent development of the discipline (Gallavotti 2006), and great challenges. A serious difficulty in the analysis of the CS is that the fluctuation-dissipation relation (see, e.g., Kubo 1966), cornerstone of quasi-equilibrium statistical mechanics, cannot be applied, because the climate is a non-equilibrium, forced and dissipative systems, where the asymptotic dynamics takes place in a strange attractor. Natural fluctuations and forced motions cannot be equivalent because, while natural fluctuations of the system are restricted to the attractor, because asymptotically there is no dynamics along the stable manifold, external forcings will induce motions out of the attractor with probability one (Ruelle 1998, 2009, Lucarini 2008, Lucarini and Sarno 2010). In a climatic context, this corresponds to an earlier intuition by Lorenz (1979) on the non equivalence between forced and free fluctuations. Therefore, it is questionable that climate change signals should project on the natural modes of variability.

Recently, Ruelle (1998, 2009) introduced a mathematical theory for computing *ab initio*, the response of a large class of non-equilibrium systems to external perturbations. The theory specifically applies only to a specific class (Axiom A) of statistical mechanical systems. Nevertheless, accepting the chaotic hypothesis (Gallavotti 2006), this class provides an excellent model for general physical systems. The applicant has further developed the Ruelle theory by considering the Fourier transform of the (linear and nonlinear) response function in the frequency domain, i.e. the susceptibility. More recently, it has been proved that Kramers-Kronig (KK) relations connect the real and imaginary part of the susceptibilities at all orders of nonlinearity. The Ruelle response theory provides a rigorous way to compute explicitly, as well-defined perturbation series, the climate response of a system to forcings featuring generic time modulation and generic spatial pattern. The KK theory and the related sum rules (Lucarini 2008b) can be used to define a comprehensive self-consistent theory of climate change against forcings of all time scales and constitute a formidable tool for assessing the consistency of a CMs, since they provide explicit and computable constraints, based only upon the principle of causality, that have to be necessarily obeyed. Models not complying with these constraints cannot feature a consistent dynamics over all of the time and space scales and require a detailed re-examination (Lucarini 2008b).

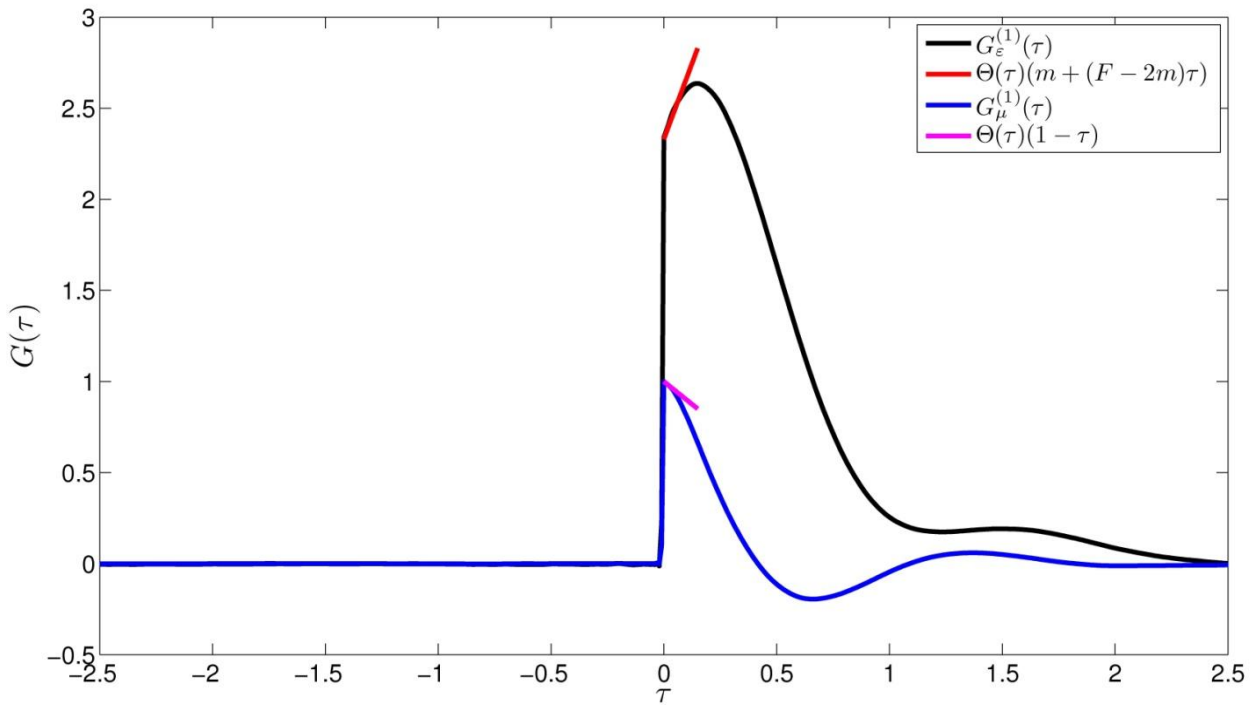

**Figure 7: Green function describing the response to a specific perturbation of the spatially averaged total energy (black line) and total momentum (blue line). The short term behaviour, computed ab initio using the response theory, is indicated in red (energy) and magenta (momentum). Adapted from Lucarini and Sarno (2010).**

The analysis of these properties with CMs of various degrees of complexity seems absolutely relevant. The prototypical numerical study by Lucarini (2009b) has been extended by Lucarini and Sarno (2010), where the first direct computation of the Green function of a simplified climate model has been performed (see Fig. 7). The Green function allows for computing ab initio the response of the considered climate observable to the external perturbation introduced into the system. Further theoretical extensions and applications to models of higher complexity and deeper climatic interest are definitely necessary.

Using the response theory formalism and its extension to the frequency domain, it is possible to compute *ab-initio* the climatic impact of quasi-static perturbations, such as those related to changes in the parameters of the system, like atmospheric composition, albedo, solar irradiation, Earth's axis inclination. Moreover, it is possible to tackle rigorously issues such as determining the impact of periodic forcings like the seasonal cycle, the solar cycle, and multi-millennial orbital variations. As in quasi-geostrophic atmospheric modelling the anomalies in topography and surface temperature appear as boundary conditions terms controlled by (small) parameters, one can compute explicitly their impact of the statistical properties of the circulation, thus extending the work of Speranza et al. (1985) on orographic modification to baroclinic instability in a climatic perspective. Similar strategy could be used for specific oceanic problems.

Finally, the analysis of the susceptibility function can highlight and quantify relevant climate feedbacks. In fact, the response of the system varies enormously with the time scale of the forcing: resonances with the internal time scales may greatly amplify the response to perturbations. On a similar note, the analysis of tipping points, as conditions under which the susceptibility diverges, could be envisioned.

6. **Conclusions**

We have briefly recapitulated some of the scientific challenges and epistemological issues related to climate science. We have discuss the formulation and testing of theories and numerical models, which, given the presence of unavoidable uncertainties in observational data, the non-repeatability of world-experiments, and the fact that relevant processes occur in a large variety of spatial and temporal scales, require a rather different approach than in other scientific contexts.

In particular, we have clarified the presence of two different levels of unavoidable uncertainties when dealing with climate models, related to the complexity and chaoticity of the system under investigation. The first is related to the imperfect knowledge of the initial conditions, the second is related to the imperfect representation of the processes of the system, which can be referred to as structural uncertainties of the model. We have discussed how Monte Carlo methods provide partial but very popular solutions to these problems. A third level of uncertainty is related to the need for a, definitely non-trivial, definition of the appropriate metrics in the process of validation of the climate models. We have highlighted the difference between metrics aimed at providing information of great relevance for the end-user from those more focused on the audit of the most important physical processes of the climate system.

It is becoming clearer and clearer that the current strategy of incremental improvements of climate models is failing to produce a qualitative change in our ability to describe the climate system, also because the gap between the simulation and the understanding of the climate system is widening (Held 2005, Lucarini 2008a). Therefore, the pursuit of a "quantum leap" in climate modelling – which definitely requires new scientific ideas rather than just faster supercomputers - is becoming more and more of a key issue in the climate community (Shukla et al. 2009). In this context, we could not disagree more with the perspective of climate science proposed in Navarra et al. (2010), who foresees a dominance of supercomputing in few selected centers, central planning of scientific priorities, and re-organisation of whole academic and scientific framework in close resemblance with what done in high-energy physics over 50 years ago. First, centralized planning of the scientific priorities (with the related allocation of funds and jobs) automatically raises the question of who is going to define such priorities and on which basis. Second, and more importantly, as widely discussed in this paper, it is hard to find to scientific sectors with as different epistemologies as high energy physics and climate science. Navarra et al. (2010) talk about "crucial experiments", but, unfortunately, these just cannot exist in a non-Galilean setting as that of climate science. In fact, the distance of climate science from the "timeless" Galilean science based upon repeated cycles of experimental investigations and improvements to scientific theory is so wide that it is impossible to apply the usual scientific validation criteria to the results of climate science. The different epistemology pertaining to climate science implies that its answers cannot be singular and deterministic, while they must be plural and stated in probabilistic terms. Flexible and open-source modelling, such as that represented by the PLASIM platform (Fraedrich et al. 2005), and distributed computing, such as that adopted in the `climaprediction.net` project (Allen 1999), seem in principle more suited for the goals, the methodologies, the development of climate science. Moreover, proposing new ideas, innovative scientific frameworks, new paradigms, rather than flexing and training metaphorical (and expensive) muscles, seem definitely more promising in the author's view.

In this regard, we have proposed the adoption of a thermodynamic perspective as a potentially relevant framework for improving our understanding of the climate system and our ability to model it. The macroscopic non-equilibrium thermodynamics allows for characterising the climate system in terms of its efficiency to produce work, i.e. organised atmospheric and oceanic motion, to achieve steady state by balancing the input and output of energy and entropy with the surrounding environment, and of its irreversibility, due to entropy-generating processes. Such global properties allow for diagnosing, characterising and understanding the smaller scale processes associated to climate variability, climate feedbacks, climate change, in general, and large scale climate re-organisations occurring at tipping points, in particular. Moreover, these tools can be used for studying the basic properties of the circulation of planetary atmospheres, a topic of great interest in the present age characterised by the discovery of quickly growing number of exoplanets.

A more fundamental approach, based upon non-equilibrium statistical mechanics, can also be envisioned. The fact that, as can be deduced from Ruelle's (1998, 2009) arguments, the climate system does not obey the fluctuation dissipation theorem, is another crucial reason why its modelling and its understanding are intrinsically difficult. The climate responses to forcings are in principle irreducible to internal fluctuations. Therefore, as opposed to common wisdom in climate science, it is not obvious at all that, e.g., climate change signals will project on natural modes of variability. Nonetheless, one should also consider that if stochastic forcing is added to the system the fluctuation-dissipation theorem is recovered (Marini Bettolo Marconi et al. 2008). Moreover, some papers have shown that a direct application of the fluctuation dissipation theorem in a climatic context is reasonably successful (see, e.g., Gritsun and Branstator 2007). It is definitely worth exploring whether this results exactly from the fact that numerical schemes introduce at all practical effects noise into the climate models, or from the fact that in the specific case of climate in present conditions the violation of the fluctuation-dissipation theorem is numerically small.

Non-equilibrium statistical mechanics provides also exciting tools for defining new strategies for the understanding of basic processes involved in large scale climate dynamics, including also feedback mechanisms, and for treating rigorously ensembles of models simulations. The Ruelle response theory and its extension in the frequency domain have been shown to allow the formulation of a new way of studying rigorously, the response of the climate system to perturbations, and to provide the foundation for defining what we may call the *spectroscopy of the climate system*, which provides the possibility of evaluating, using a perturbative approach, climate sensitivity and climate change from a radically new perspective. This paves the way for studying a potentially immense class of problems.

*Acknowledgements*
*VL wishes to thank GGG for providing him food for thought.*